  \documentclass[12pt]{article}

  \usepackage{graphicx}
  \usepackage{amssymb}
  \usepackage{textcomp}
  \usepackage{endnotes}
  \usepackage{xcolor}
  \usepackage[colorlinks=true,urlcolor=blue]{hyperref}
  \usepackage{lineno}
  \usepackage{colortbl}
  \usepackage{amsmath} 
  \usepackage{comment} 

  \usepackage[font={footnotesize,sf,bf}]{caption}
  \usepackage{soul}

\begin{document}

  \let\footnote=\endnote
  \renewcommand{\notesname}{\center \normalsize \bf Endnotes}

  \DeclareGraphicsExtensions{.pdf,.jpg} 

  \def\keff{$k_{\mbox{\scriptsize eff}}~$}

  \setlength{\fboxsep}{0pt}
  \setlength{\fboxrule}{0.5pt}


\begin{center} {\Large

Estimating Potential Tritium and Plutonium Production \\
in North Korea's Experimental Light Water Reactor

}

\medskip

Patrick J. Park and Alexander Glaser \\[0.5ex] {\footnotesize \em Program on Science and Global Security, Princeton University}

\end{center}


\begin{quote} \small {\bf \bf Abstract.} Our work explores North Korea's 100 MW-th Experimental Light Water Reactor (ELWR) and its potential contributions to the country's nuclear weapons program. Built at the Yongbyon Nuclear Research Center, the ELWR began operations in October 2023 and represents North Korea's first attempts at a light-water reactor using domestically-enriched, ceramic fuel. Our study examines possible configurations for energy, tritium, and tritium-plutonium co-production. Assuming a single-batch core, the ELWR can be used to produce 48--82 grams of tritium, which can supply 2--4 new boosted warheads each year, up to a maximum arsenal of 88--150 warheads total. Concurrent production of tritium and weapon-grade plutonium is also possible but requires reprocessing of spent ceramic fuel. These findings underscore how North Korea's nuclear capabilities may be advanced through the ELWR's dual-use potential. \end{quote}

Correspondence: Patrick Park (\href{mailto:pp0191@princeton.edu}{pp0191@princeton.edu}), Program on Science and Global Security, Princeton University, 221 Nassau St, 2nd Floor, Princeton, NJ 08542 (USA).


  \bigskip

{\centering \bf Background\par}%

\begin{figure}[!bht]
  \begin{center} \sf 
  \fbox{\includegraphics[height=42mm]{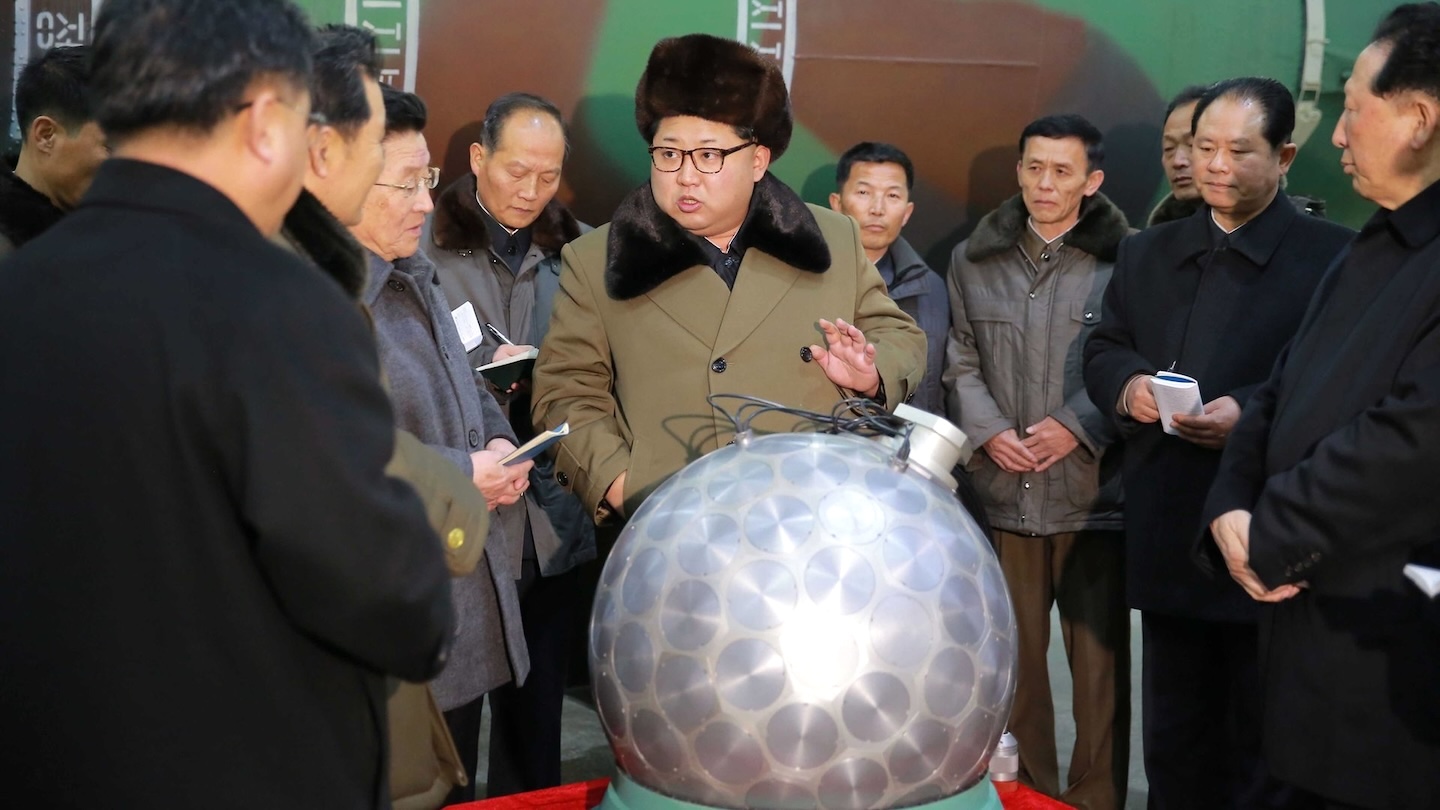}}\hspace{1ex}%
  \fbox{\includegraphics[height=42mm]{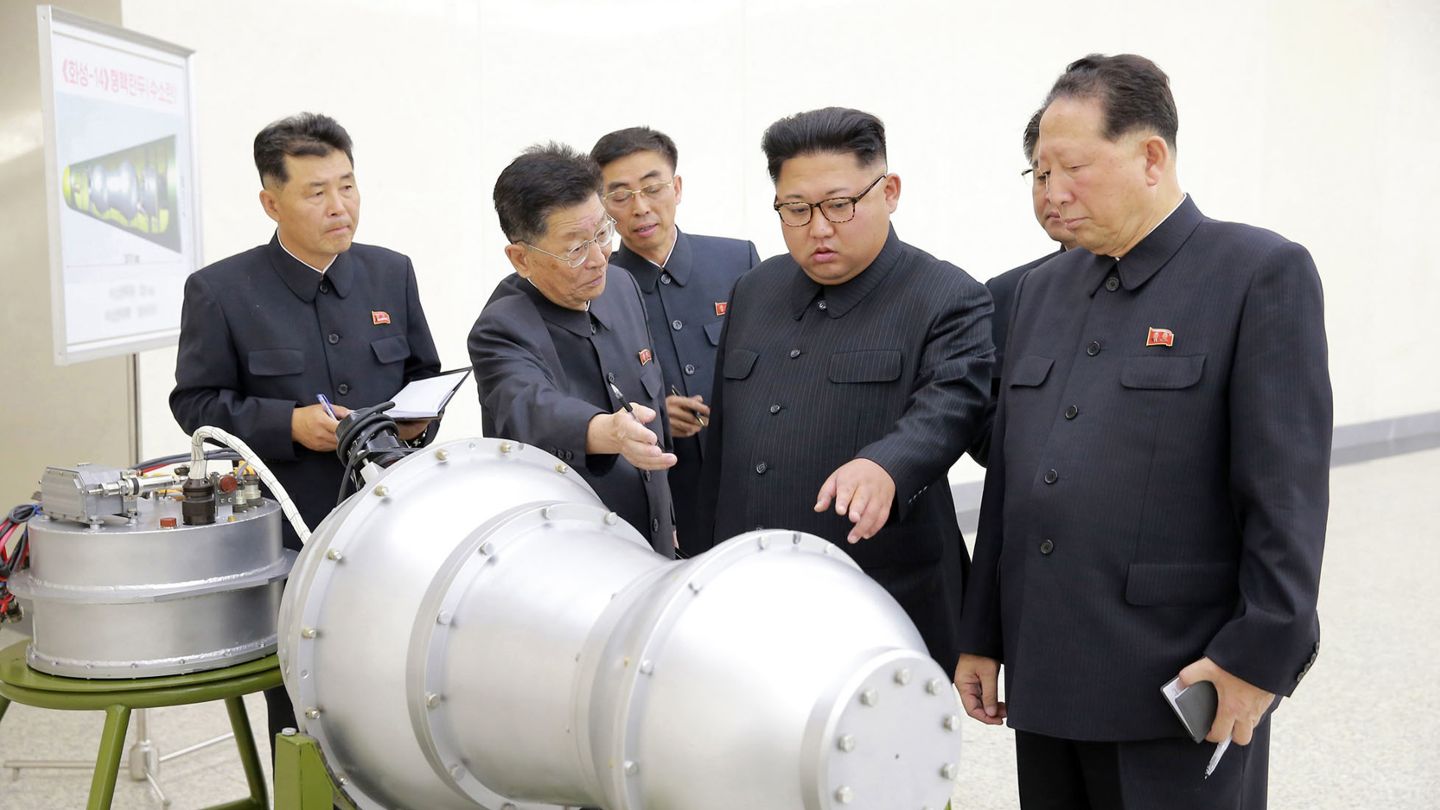}}
  \caption{North Korean nuclear weapons. {\sf In March 2016, North Korea first published photos of a probable nuclear weapon, shown on the left (``the disco ball''); that year, it conducted two nuclear tests, both of which have been considered successful. On September 2, 2017, North Korea published a photo of a two-stage weapon, shown on the right (``the peanut''). The following day, it conducted a large nuclear weapon test with an estimated yield of 250 kt(TNT). The diameter of the disco ball is consistent with the size of the primary (on the left-hand side in the photo) of the two-stage weapon; both designs suggest a boosted primary, which requires the injection of a deuterium-tritium gas prior to detonation of the device.} {\sl Source: Korean Central News Agency.}}
  \label{fig:weapons}
  \end{center}
  \end{figure}

North Korea's elusive nuclear program has advanced rapidly from its first sub-kiloton and possibly failed nuclear test in 2006 to the detonation of a two-stage thermonuclear weapon in 2017 (Figure~\ref{fig:weapons}). As of 2024, North Korea is estimated to have assembled on the order of fifty nuclear weapons,\footnote{Hans M. Kristensen, Matt Korda, Eliana Johns, and Mackenzie Knight, \href{https://doi.org/10.1080/00963402.2024.2365013}{North Korean Nuclear Weapons, 2024}, {\em Bulletin of the Atomic Scientists,} 80 (4), 2024.} but Kim Jong Un has declared the goal to ``exponentially expand'' North Korea's nuclear arsenal in coming years.\footnote{``The present situation highlights the importance and necessity of mass-producing tactical nuclear weapons and demands an exponential increase in the country's nuclear arsenal... [this being] the epochal strategy of the development of nuclear force and national defense for 2023,'' \href{http://rodong.rep.kp/en/index.php?MTJAMjAyMy0wMS0wMS1IMDA1QDE1QDFAQDBAMQ==}{Report on the 6th Enlarged Plenary Meeting of 8th WPK Central Committee}, {\em Rodong Sinmun}, January 1, 2023.}


There is some evidence that, from the outset, North Korea sought to use fissile materials as efficiently as possible, and tritium-boosted weapons may be one strategy to achieve this goal.\footnote{The process of boosting can be summarized as follows: ``The high-energy (14 MeV) neutrons liberated in the D-T reaction are used in many fission weapons to achieve what is known as `boosting.' Neutrons from the D-T reaction are introduced at a late stage of the fission chain in order to maintain and enhance the progress of the fission reactions. There is a considerable increase in the energy released because of the greatly improved efficiency in utilization of the fissile material.'' Excerpt from \S1.49 in Samuel Glasstone and Leslie M. Redman, {\em An Introduction to Nuclear Weapons,} WASH-1034 (revised), U.S. Atomic Energy Commission, Washington, DC, 1972, \href{https://ipfmlibrary.org/aec72.pdf}{ipfmlibrary.org/aec72.pdf}.} This is supported by its Nuclear Weapons Institute’s claim of having successfully tested ``rapidly boosting fission-fusion reactions'' in a two-stage weapon in 2017.\footnote{``Symmetrical compression of nuclear charge, its fission detonation and high-temperature nuclear fusion ignition, and the ensuing rapidly boosting fission-fusion reactions, which are key technologies for enhancing the nuclear fusion power of the second-system of the H-bomb, were confirmed to have been realized on a high level,'' \href{https://kcnawatch.org/newstream/1529087210-172788751/dprk-nuclear-weapons-institute-on-successful-test-of-h-bomb-for-icbm/}{DPRK Nuclear Weapons Institute on Successful Test of H-bomb for ICBM}, {\em Ryugyong}, September 6, 2017.} It has not been clear, however, how North Korea could produce tritium in sufficient quantities to support its nuclear arsenal.\footnote{Existing analyses have often focused on the 5 MWe graphite-moderated reactor, but tritium production potential of this facility is rather limited; see for example, Sungmin Yang et al., ``Estimating North Korea's Nuclear Capabilities: Insights From a Study on Tritium Production in a 5MWe Graphite-moderated Reactor,'' {\em Nuclear Engineering and Technology,} 56, 2024; Siegfried S. Hecker, Chaim Braun, Christopher Lawrence, Panos Papadiamantis, \href{https://fsi.stanford.edu/publication/north-korean-nuclear-facilities-after-agreed-framework}{North Korean Nuclear Facilities After the Agreed Framework}, {\em Stanford FSI Working Papers}, July 1, 2016.}

\begin{figure}[hbt]
  \begin{center} \sf 
  \includegraphics[height=55mm]{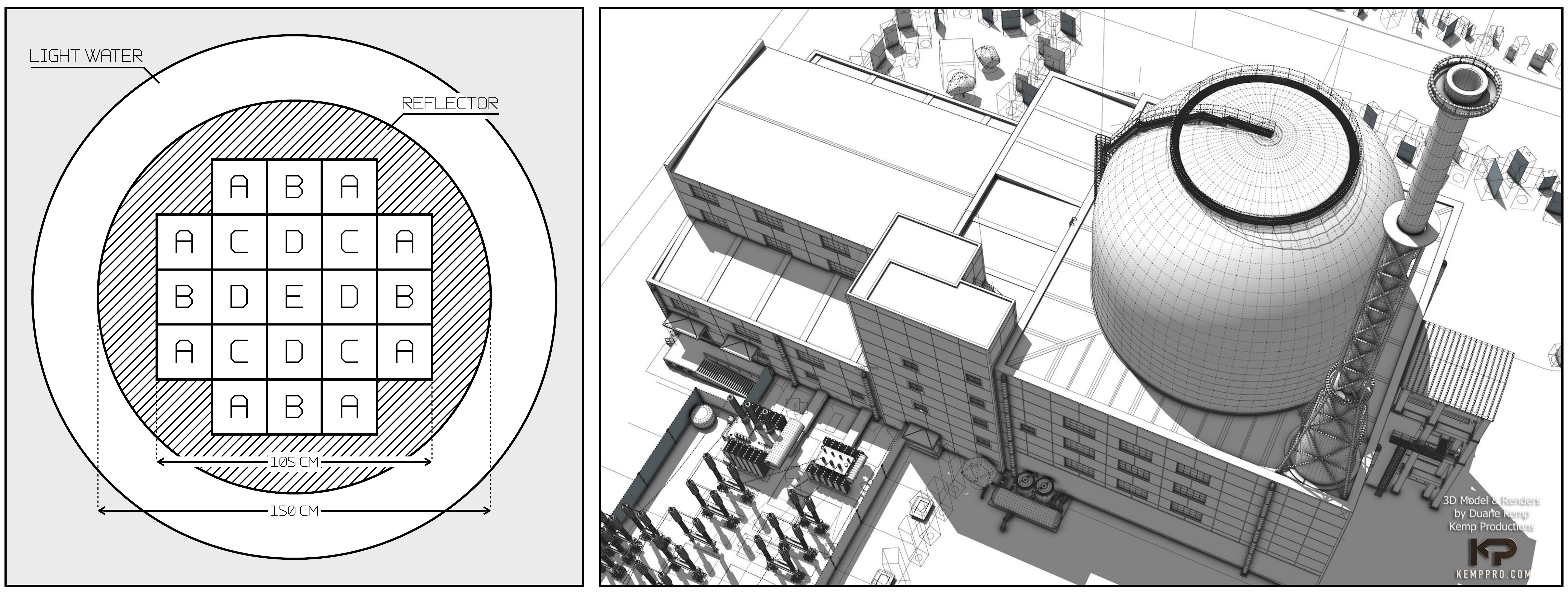} \caption{Core configuration used for all full-core depletion calculations. \sf{Based on the locations in the core, isotopics are tracked for five different types of fuel assemblies (A--E), each with two depletion zones, i.e., a total of ten zones for the reactor core. Shown on the right is a rendering of the ELWR with some of the associated support facilities, including the electrical switchyard. The reactor dome has a diameter of about 23 meters, comparable to a small research reactor. \textit{3D rendering by Duane Kemp, \href{https://www.kemppro.com}{kemppro.com}.}}
  }
  \label{fig:model}
  \end{center}
  \end{figure}

In October 2023, observation of hot effluent release in satellite imagery confirmed the operation of the new Experimental Light Water Reactor (ELWR) at (\href{https://maps.app.goo.gl/hPLqih7henBZZZm56}{39.79587$\:$N, 125.75507$\:$E}). It is their fourth nuclear reactor and the third constructed at the Yongbyon Nuclear Research Center, the heart of North Korea's nuclear program located 100~km north of Pyongyang, along the Kuryong River. The ELWR has a nominal power of 100~MW thermal and reportedly uses uranium dioxide fuel enriched between 2.2\% to 4.0\%, for an average of 3.5\%, marking the country's first attempt at producing ceramic and domestically-enriched fuel.\footnote{Siegfried S. Hecker, \href{https://cisac.fsi.stanford.edu/publications/north_koreas_yongbyon_nuclear_complex_a_report_by_siegfried_s_hecker}{A Return Trip to North Korea’s Yongbyon Nuclear Complex}, {\em Stanford CISAC Policy Briefs}, November 20, 2010. Though North Korea actually claimed to produce fuel enriched between 2.2--4.0\% for the ELWR, we did not examine more complex core loading patterns (inside-out, checkerboard, low-leakage, etc.) because these require several additional modeling assumptions. Given the lack of information, we applied a uniform enrichment of 3.5\% in our ELWR model, which should have only minor effects on the main results.} State representatives claimed the ELWR is a testbed for future commercial LWRs and to supply local power, evidenced by its turbine hall, switchyard, and transmission lines---the first such setup for a reactor in the country (Figure~\ref{fig:model}, right).\footnote{Sulgiye Park and Allison Puccioni, \href{https://www.38north.org/2024/01/north-koreas-pursuit-of-an-elwr-potential-power-in-nuclear-ambitions/}{North Korea’s Pursuit of an ELWR: Potential Power in Nuclear Ambitions?}, {\em 38 North}, January 24, 2024.} 

North Korea's primary producer of weapons plutonium is believed to be the 5~MWe ``Yongbyon Test Reactor No. 1'' at up to 6~kg or about one weapon's worth per year.\footnote{``Estimating Plutonium Production in North Korea,'' Appendix 3B in {\em Global Fissile Material Report 2009: A Path to Nuclear Disarmament,} International Panel on Fissile Materials, Princeton, NJ, October 2009.} As this limited output may constrain a rapid arsenal growth, prior studies have explored the ELWR's potential for plutonium production.\footnote{Cecilia Gustavsson, Peter Andersson, Erik Branger, Grant Christopher, David Schmerler, and Hailey Wingo, ``Modelling Fissile Production in the Experimental Light Water Reactor (ELWR) of DPRK,'' Alva Myrdal Centre for Nuclear Disarmament, Annual Conference, Uppsala University, Sweden, June 2024; Seungnam Lee, Beom Jin Kim, and Ser Gi Hong, ``Preliminary Estimates of Nuclear Weapon Potential in North Korea's New ELWR,'' {\em Transactions of the Korean Nuclear Society Spring Meeting,} Jeju, Korea, May 9--10, 2024.} We also consider this scenario below, but it may require North Korea to acquire the capability to reprocess ceramic fuel, which is technically more demanding than separating plutonium from the metal fuels it has processed so far.

Given North Korea's fissile material constraints, tritium-boosted weapons emerge as an attractive option, allowing the construction of more weapons with similar yield using much less fissile material per device. While the existing Yongbyon reactor may have produced the tritium used in the 2017 test, its low power and low reactivity margin limit sustained tritium production compared to the ELWR. This paper develops a neutronics model of the ELWR based on the best available information to estimate possible tritium and plutonium production rates and related fuel-cycle characteristics. These estimates are intended to provide an upper limit on North Korea's potential arsenal expansion.

\bigskip

{\centering \bf Reactor Model and Neutronics Calculations\par}%

The results presented in this paper are based on full-core reactor calculations carried out with the MCODE computer code system,\footnote{Xu, 2003, {\em op.$\,$cit.} We use MCODE Version 1.0 with minor changes in the source code so that MCNP6 tally files can be parsed correctly. Note that there are more recent releases of MCODE (2.2 and 3.0), but we do not need the added functionalities implemented in those versions.} which links the Monte Carlo neutron transport code MCNP6 with the ORIGEN2 point-depletion code and permits reactor burnup calculations with regularly updated neutron flux distributions and one-group cross sections.\footnote{Scott B. Ludwig, Revision to ORIGEN2, Version 2.2. Oak Ridge National Laboratory, May 2002. Allen G. Croff, A User's Manual for the ORIGEN2 Computer Code. ORNL/TM7175, Oak Ridge National Laboratory, July 1980, Allen G. Croff, \href{https://www.tandfonline.com/doi/abs/10.13182/NT83-1}{ORIGEN2: A Versatile Computer Code for Calculating the Nuclide Compositions and Characteristics of Nuclear Materials}, {\em Nuclear Technology,} 62 (3), September 1983.} MCODE has been extensively benchmarked against other depletion codes.


We based our core design around a standard Westinghouse-style 17$\times$17 fuel assembly, which is also the basis of South Korea's OPR1000 and APR1400 reactor fuel assemblies. The ELWR has been reported to have a core inventory of 4,000~kg of UO$_2$, and we use this amount as our reference value for all core variants discussed below. We model the core as a 5$\times$5 grid of 21 fuel assemblies, giving it an active height of 144~cm and a maximum width of 105~cm (Figure~\ref{fig:model}, left). This geometry is close to the optimal height-to-width ratio for cylindrical cores and results in a power density that is typical for pressurized-water reactors. Each assembly is coded to contain two burnup zones. Modeling five distinct assembly types, as shown in Figure~\ref{fig:model}, yields a total of 10 burnup zones in the core. In our model, the UO$_2$ was uniformly enriched to 3.5 wt\% U-235. Fuel pins could be clad in either zircaloy or stainless steel. Zirconium alloys are commonly used in power reactors but are difficult and expensive to fabricate. Stainless steel is cheaper and has more favorable mechanical properties, but it significantly reduces the discharge burnup due to the larger neutron capture cross section of iron-56.

To examine the relevance of the choice of cladding material, we modeled both options: zircaloy-4 cladding provides an upper bound for the performance of the ELWR, while 304 stainless-steel cladding represents a more conservative choice, especially during initial commissioning. To offset the penalty of stainless steel, we assumed that the thickness of the cladding is reduced to 0.400~mm compared to the reference value of 0.573~mm used for zircaloy.\footnote{For reference, early American power reactors like Connecticut Yankee used 304 stainless-steel cladding of thickness 0.419 mm before switching to zircaloy; see Table~1 in V. Pasupathi and R. W. Klingensmith, \href{https://digital.library.unt.edu/ark:/67531/metadc1053016/m2/1/high_res_d/5160258.pdf}{\em Investigation of Stainless Steel Clad Fuel Rod Failures and Fuel Performance in the Connecticut Yankee Reactor}, EPRI NP-2119, Project 1758-1, Battelle Laboratories, Columbus, OH, November 1981.}

A core inventory of four metric tons of UO$_2$, regardless of the particular assembly design, yields a relatively small core. To minimize neutron leakage, we assume the presence of a thick radial and axial reflector, which increases the core non-leakage factor from about 0.93 to 0.96. Beryllium oxide and graphite reflectors are both viable options, but the choice of either material does not significantly affect reactor core neutronics.

In all scenarios examined below, we assume that the reactor is operated in a single-batch mode, i.e., the entire core is discharged and reloaded after each cycle. This simplifies core management considerations for the operators. Given the relatively small number of fuel assemblies in the core, the advantages of multi-batch cores may be considered marginal. In fact, single-batch cores maximize the cycle length, reduce refueling outages, and significantly improve the capacity factor of a reactor, albeit at the expense of uranium throughput and enrichment requirements.\footnote{See Figure 2.11, in Zhiwen Xu, \href{https://dspace.mit.edu/handle/1721.1/16603}{Design Strategies for Optimizing High Burnup Fuel in Pressurized Water Reactors}, PhD thesis, Massachusetts Institute of Technology, January 2003.} However, low fuel burnup is required if the production of weapon-grade plutonium is desired, which provides another rationale for using a single-batch core.

To explore various end uses of the ELWR, we analyze three different core configurations. First, we establish a baseline ``clean core'' that simply maximizes discharge burnup; this represents a civilian usage where electricity production is prioritized. Second, we consider a ``tritium core'' in which lithium targets are introduced in the core in a way that maximizes the tritium production per cycle. These first two cores do not yield weapon-grade plutonium. Third, we examine a ``co-production core,'' where the lithium loading is increased such that the core reaches its end-of-life just when the plutonium is still weapon-grade (90~wt\% Pu-239). This arrangement produces tritium and weapon-grade plutonium at the same time, but the discharge burnup is much lower and the reactor cycle therefore also shorter. Combined with our two cladding variants (zircaloy vs. stainless steel), we present results for a total of six (2$\times$3) cores.

For the tritium core, we devised an iterative procedure that finds the optimum lithium loading, balancing reaction rates and cycle length, to maximize the total tritium produced by the end of the cycle. Since MCODE cannot directly deplete lithium, we defined an offline constrained convex optimization problem to obtain these results.

\begin{figure}[hbt]
    \centering
    \includegraphics[width=130mm]{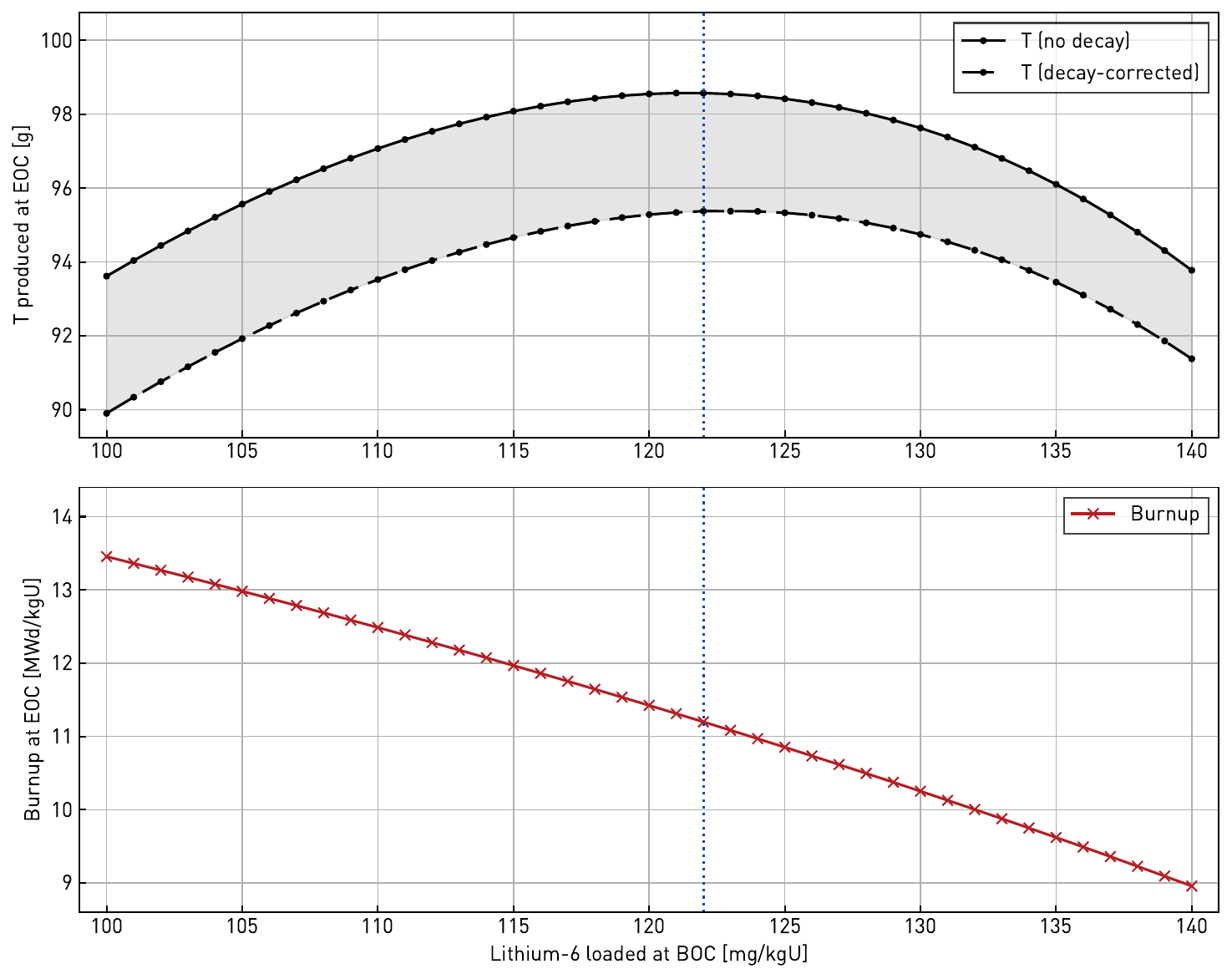}
    \caption{Total tritium production versus lithium loading. {\sf As the amount of lithium in the core increases, the instantaneous tritium production rate also increases; at the same time, however, the achievable discharge burnup decreases, which reduces the total amount of time that the reactor operates and tritium can be produced. Clearly, for a given core configuration, a lithium loading must exist that maximizes total tritium production during the cycle. In this example, an initial lithium loading of 122 mg/kg results in a discharge burnup of 11.2~MWd/kg; at that point, a total of 98.6~grams of tritium have been produced, of which 95.4~grams still exist upon discharge of the targets. Numerical results are obtained by solving a convex optimization problem using effective cross section and core reactivity values determined in MCNP6.}}
    \label{fig:optimization}
\end{figure}

As the presence of lithium-6 introduces negative reactivity, loading too much lithium leaves the core subcritical before an optimal amount of tritium is produced; in contrast, too little lithium does not take advantage of the available reactivity margin. We model this trade-off by fitting a quadratic ``$k$-penalty'' function, mapping lithium concentration to negative reactivity with numerical data from MCNP6 pin-cell calculations. Using this penalty, reactor $k_\text{eff}$ is recursively docked as the lithium is depleted. The cycle ends when $k_\text{eff} = 1.03$. Given these constraints, the optimization problem can be solved iteratively to obtain the optimal lithium loading. Figure~\ref{fig:optimization} illustrates the procedure.

\bigskip

{\centering \bf Results\par}%

\begin{figure}[hbt]
    \centering
    \includegraphics[width=120mm]{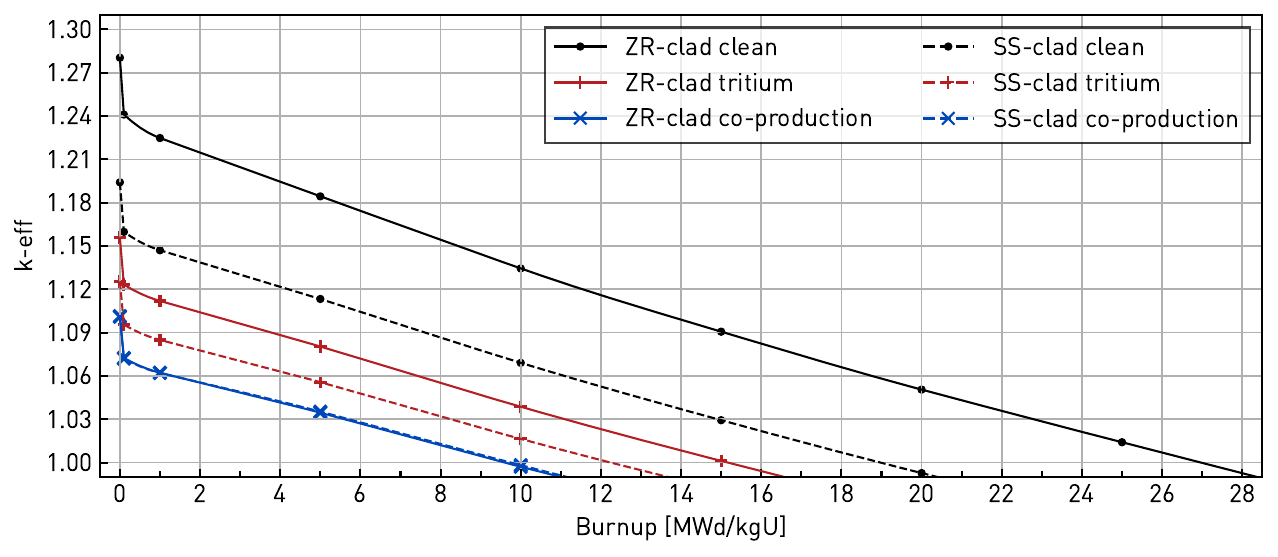}
    \caption{Core reactivity \keff vs fuel burnup. {\sf We assume that the core reaches its end of life once \keff drops below 1.03. The discharge burnup strongly depends on the type of cladding used and the amount of lithium-6 present in the core. All results are from full-core MCODE6 simulations.}}
    \label{fig:keff-ss}
\end{figure}

\begin{figure}[hbt]
    \centering
    \includegraphics[width=120mm]{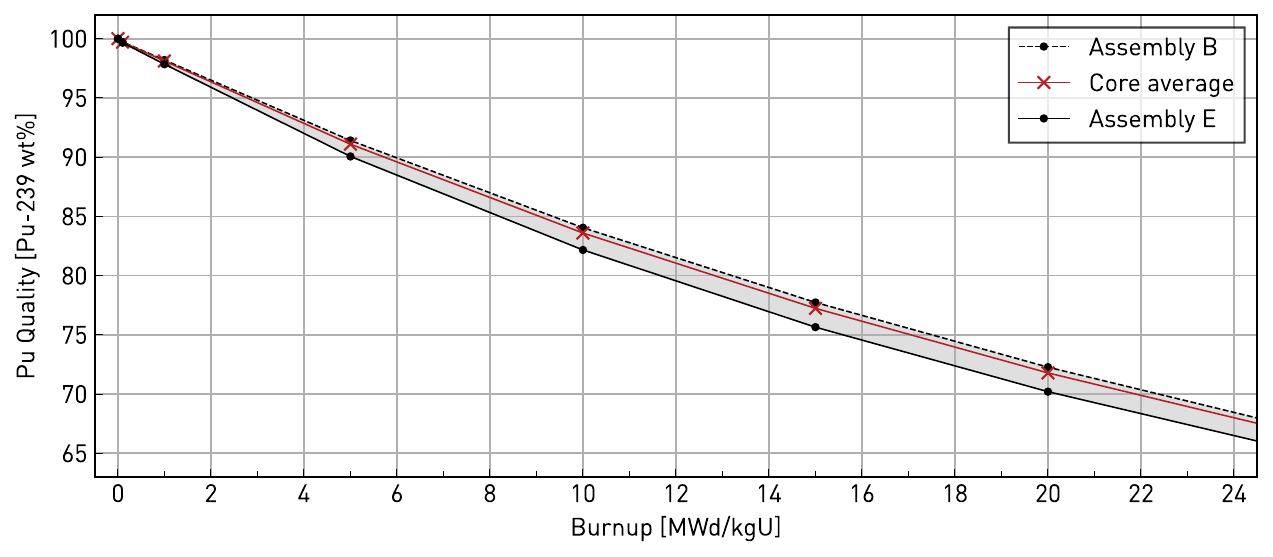}
    \includegraphics[width=120mm]{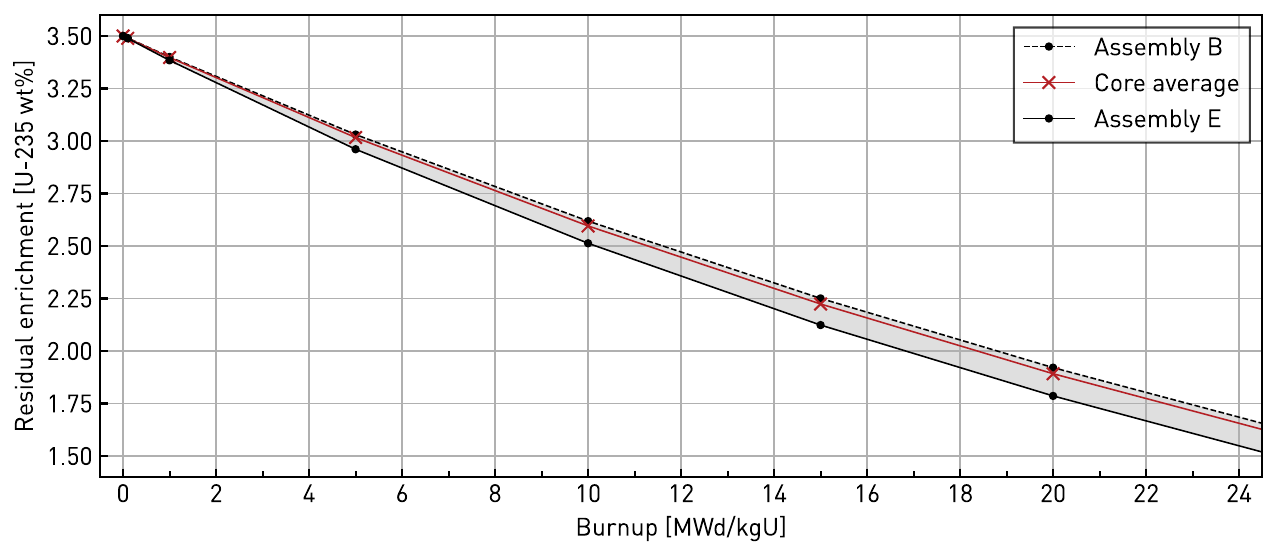} \\
    \caption{Plutonium-239 fraction and residual uranium-235 enrichment versus fuel burnup. {\sf The evolution of these values are virtually identical for all cases considered; shown here are results for the clean core. Shaded areas indicate variations depending on the position of the fuel assemblies in the core. All results are from full-core MCODE simulations.}}
    \label{fig:isotopics}
\end{figure}

The main results for all six core variants are summarized in Tables~\ref{tab:zircaloy} and \ref{tab:stainless} below. The achievable discharge burnup strongly depends on the type of cladding and on whether tritium production is pursued. Figure~\ref{fig:keff-ss} summarizes the reactivity coefficient \keff versus burnup. In all cases, we assume that end-of-cycle is reached when \keff has dropped to 1.03. The evolution of plutonium isotopics and residual uranium-235 enrichment are virtually identical for all cases considered (Figure~\ref{fig:isotopics}). Separative work requirements were estimated using the standard formula for uranium enrichment assuming a tails depletion level of 0.27\%.\footnote{A North Korean centrifuge engineer reported an average product enrichment level of 3.5\% and a tails depletion level of 0.27\%; see p.~4 in Hecker, 2010, \textit{op.$\,$cit}.} The discussion below highlights the main findings for each of the cores considered.

{\centering \em Clean Core\par}%

The ELWR achieves a relatively low discharge burnup of 14.9--22.8 MWd/kg, primarily due to its smaller core size. Stainless-steel cladding was found to shorten cycle length by about 35\% compared to zircaloy cladding (526 days versus 804 days at full power). Despite this limitation, the ELWR demonstrates the potential to operate continuously at full power with a cycle length for at least 17~months, comparable to current commercial light-water reactors. 

When operated in this mode aimed at maximum discharge burnup, the reactor could in principle sustain regular operations at a capacity factor of more than 94\% if the refueling can be completed within 30 days. Annual requirements of separative work are between 7,110 SWU/yr and 10,700 SWU/yr depending on the type of cladding used. While both values may well be within North Korea's estimated enrichment capacity, this range also emphasizes the importance of fuel design information that remains unknown to outside analysts at this time. Plutonium production is about 11.8--13.6~kg/yr, but the isotopics may be considered unfavorable for use in weapons.\footnote{Weapon designs prefer a plutonium-239 percentage of 90\% or more, even though lower grade plutonium is also weapon-usable; see J. Carson Mark, ``Explosive Properties of Reactor-grade Plutonium,'' {\em Science \& Global Security,} 4, 1993.}


\medskip

{\centering \em Tritium Core \par}%

We consider the possible production of tritium in the ELWR one of the most ``attractive'' features that this facility may offer to support North Korea's nuclear weapons program. In order to estimate tritium production in the ELWR, we add small amounts of lithium to the cladding material, though a practical target design may be different.\footnote{Embedding lithium in the cladding simplifies the analysis, without having to make further assumptions about the target design. While the actual location and distribution of the material are likely to be different, tritium production rates and core reactivity penalties are largely determined by the total lithium loading and respective net neutron absorption rates. Other design details are far less relevant.} Typical values are on the order of 100 milligrams of lithium-6 per kilogram of uranium in the fuel. In our simulations, we used material that is enriched to 90\% in the isotope lithium-6, but natural lithium could equally be used as long as an equivalent amount of lithium-6 is present.\footnote{Lithium enrichment is a well-established process, but it requires large amounts of mercury, making it environmentally highly problematic. Both lithium-6 and lithium-7 have industrial applications. China and Russia are the main suppliers of enriched lithium today. North Korea is believed to have developed industrial-scale lithium enrichment capabilities following a UN investigation into its attempted online sale of lithium-6 metal in 2016; see \S IV.A.3.25, \href{http://undocs.org/S/2017/150}{Report of the Panel of Experts Established Pursuant to Resolution 1874 (2009)}, {\em UN Security Council Committee Report S/2017/150}, February 17, 2017.} Our total production estimates listed in the tables below also account for concurrent tritium decay during production, a relatively small but noticeable effect.

Overall, we find that tritium production is maximized when the cycle length is roughly reduced by a factor of two compared to the clean core. For the core using fuel pins with zircaloy cladding, the new cycle length is 393 days. During this cycle, 95.4 grams of tritium is produced; this translates to a production rate of 82.2 grams per year (Table~\ref{tab:zircaloy}). Performance of the stainless-steel core is significantly inferior: in this case, only about 48.3 grams of tritium are produced per year (Table~\ref{tab:stainless}).

It is worth noting that plutonium production per year (but not per cycle) is higher than for the clean core and reaches 14.7--15.1~kg/yr; this is due to the lower burnup of the fuel. As a result, the fraction of Pu-239 is also higher (82.3--86.2\%), but the material is still not considered weapon-grade.

The reduction in cycle length corresponds to a similar increase in the requirements for the supporting fuel cycle, i.e., both the demand of natural uranium and separative work roughly increase by a factor of two compared to the clean core. The residual enrichment of the fuel, however, is significantly higher (2.52--2.75\%), and the operator might not want to discard this material without further use, as we will discuss later.

\medskip

{\centering \em Co-Production of Tritium and Weapon-Grade Plutonium \par}%

A third approach we consider is the co-production of tritium and weapon-grade plutonium, which adds a constraint for the maximum discharge burnup. MCNP6 simulations show that the ELWR can only operate for 200~days at nominal power before the plutonium-239 content drops to 90\%. Any further operation renders the plutonium below weapon-grade.\footnote{North Korea appears to prefer even super-grade plutonium for their weapons. During the Stanford team's 2010 visit to North Korea, Yongbyon Director Ri Hong Sop was quoted as considering fuel with burnup exceeding 3~MWd/kgU as unusable for weapons; this burnup corresponds to plutonium-239 fraction of about 95\%.} This exposure corresponds to a discharge burnup of only 5.67~MWd/kgU. Given that the target burnup is so low, the lithium loading can be further increased such that the core reaches its end of life, i.e., \keff $\approx$ 1.03, at that target burnup.

As shown in Tables~\ref{tab:zircaloy} and \ref{tab:stainless}, annual production of weapon-grade plutonium is on the order of 15.5~kg/yr, irrespective of the cladding material. In contrast, and as in the previous case, tritium production strongly depends on the type of cladding: it reaches 115.6 grams per year for the zircaloy core and 56.0 grams per year for the stainless-steel core, i.e., it is even higher than the annual production rates obtained for the tritium core. Tritium production per cycle is however 16--24\% lower than for the dedicated tritium core. Due to the low discharge burnup and the high lithium loading, the co-production core requires significantly more enriched lithium and other resources compared to the tritium core.

\bigskip

{\centering \bf Discussion \par}%

The ELWR appears to offer the first robust source of tritium for North Korea's nuclear weapons program. If operated with adequately designed lithium targets, the reactor could produce 48--82 grams of tritium per year.

Having a steady supply of tritium could be particularly important for a weapons program that is constrained by the availability of fissile material. If the efficiency of the fission chain reaction in a primary can be increased by boosting, then less material is needed per warhead and more warheads can be made with the same amount of material. More importantly perhaps, boosted warheads can be lighter than an unboosted warhead of a similar yield, which increases the range of the ballistic missile it is deployed on. All these factors may be considered relevant for North Korea's evolving nuclear weapons program.

A supply of tritium can be used in two ways: it can provide the material needed for new weapons, i.e., to support a growing arsenal, or it can be used to replenish tritium that has decayed in an existing stockpile of warheads. Tritium decays at a rate of 5.47\% per year.\footnote{The half life of tritium is 12.32 years; after one year, $\exp[-\ln(2)/12.32] \approx 0.9453$, i.e., 94.53\% of the original material remains.} If we assume that 10 grams of tritium are required or allocated per warhead,\footnote{The United States has declassified the fact that ``the amount of tritium in a reservoir is typically less than 20 gm,'' {\em Restricted Data Declassification Decisions 1946 to the Present, RDD-8,} U.S. Department of Energy, Washington, DC, January 2002. Other U.S. Government reports suggest that about 7 grams of tritium are allocated for each warhead in the stockpile. This estimate is based on the requirement to make 1,500 grams of tritium per year to sustain the START II stockpile, which could have been on the order of 4,000 warheads; see {\em Tritium Production Technology Options,} Defense Science Board, U.S. Department of Defense, Washington, DC, 1999.} then 0.547 grams are lost per device and year. {Annually producing 48--82 grams of tritium can make up for this loss in 88--150 boosted warheads. Alternatively, the supply could be used to grow North Korea's nuclear arsenal. If we arbitrarily assume that about half the supply can be allocated to new weapons while existing weapons are being replenished, then North Korea's arsenal of boosted weapons could grow at a rate of about 2--4 warheads per year. In practice, of course, North Korea could combine both strategies and gradually grow its arsenal, until a plateau is reached. The material supplied by the ELWR can sustain a stockpile of no more than the 88--150 warheads estimated here.

To operate the ELWR at nominal power, North Korea needs a supply of low-enriched uranium. For this analysis, we assumed that the core uses 3.5\%-enriched material throughout the core. At a minimum, this would require 7,100~SWU per year if the reactor uses zircaloy cladding and is operated as a ``clean core,'' i.e., in civilian mode without tritium and weapon-grade plutonium production. In this case, the core reaches an average discharge burnup of 22.8~MWd/kg and about 11.3~metric tons of natural uranium are needed annually to make the fuel for the reactor (Table~\ref{tab:zircaloy}, left column). Stainless steel cladding reduces the discharge burnup by more than 30\%, leading to a respective increase in natural uranium demand to 17.0 tons and a separative work of 10,700 SWU/yr (Table~\ref{tab:stainless}, left column).

$$ \star $$

For the tritium core, in which the lithium loading is optimized for maximum tritium production per cycle, additional resources are needed. As summarized in Tables~\ref{tab:zircaloy} and \ref{tab:stainless} (center columns), demands for natural uranium and separative work essentially double compared to their baseline values for the two claddings considered (zircaloy and stainless steel). This would raise the separative work requirements to 14,000--18,500~SWU per year, which may still be within North Korea's available enrichment capacity, especially when considering the most recent revelations about a second centrifuge enrichment plant.\footnote{Ju-min Park and Josh Smith, ``Photos Likely Show Undeclared North Korea Uranium Enrichment Site, Analysts Say,'' {\em Reuters,} September 15, 2024.} It does raise the question, however, of whether North Korea would dedicate such a large fraction of its enrichment capacity to the production of low-enriched uranium instead of making weapon-grade highly enriched uranium.

As discussed above, in the tritium core, the residual uranium enrichment of the spent fuel is much higher (2.52--2.75\%) due to the lower discharge burnup. In principle, the operator could recover and re-enrich this uranium, which would drastically reduce the resources required for a new core. Since the plutonium would not be weapon-grade (82.2--86.2\% Pu-239, Figure~4), however, it is not clear whether such reprocessing campaigns would be considered worthwhile. Alternatively, the lithium targets could be removed from the core when the discharge burnup is reached, switching to ``civilian mode'' for the rest of the reactor cycle. This would effectively double the life of the core. Of course, this would also decrease net tritium production by a factor of two as the reactor would not be producing tritium half of the time.

$$ \star $$

Finally, North Korea could seek to make weapon-grade plutonium in the ELWR. As shown Figure 4, this limits the discharge burnup to at most about 5--6~MWd/kg, which is much lower than what the reactor could otherwise reach when fueled with 3.5\%-enriched uranium. Plutonium production could be as high as 15~kg per year. This represents a significant increase in production capacity when compared to the rate offered by the original Test Reactor No.1 at Yongbyon. In the case of the zircaloy core, tritium production further increases to 115.6 grams per year. Still, co-production of plutonium and tritium appears as a rather inefficient strategy unless the material in spent fuel is systematically reused. 

In our simulations, a discharge burnup of 5.67~MWd/kg is reached after 200 effective full power days. At this point, the fuel still has a uranium-235 content of about 3.0\%. Assuming that the entire core is discharged and reloaded with fresh fuel, then the annual demand of 3.5\%-enriched uranium is on the order of 5,600~kg (Tables~2 and 3, right column). This would require about 25,000~SWU/yr if natural uranium is used as feedstock. A more plausible strategy could envision the re-use of pre-enriched uranium recovered during reprocessing. For example, it would take only about 32~kg of weapon-grade HEU to ``up-blend'' 5,600~kg of uranium to its original enrichment level of 3.5\%.\footnote{Here, we solve: $5600 \mbox{ kg} \times 0.035 = (5600 \mbox{ kg} - x) \times 0.030 + 0.9 \, x$, where $x$ is the amount of blendstock needed; in other words, $x \approx 32.2$ kg of weapon-grade uranium enriched to 90\% are sufficient to raise the enrichment level of the uranium recovered from the spent fuel from 3.0\% to 3.5\%.} Alternatively, the uranium recovered from the spent fuel could be re-enriched:
\[
P = 5600 \mbox{ kg} \times \bigg( \frac{0.030 - 0.007}{0.035 - 0.007} \bigg) = 4600 \mbox{ kg}
\]

In other words, more than 80\% of the fuel could be refabricated using the uranium recovered from a previous core. The separative work required for this process is rather small:
\[
\delta U = 4600 \mbox{ kg} \times V(0.035) + 1000 \mbox{ kg} \times V(0.007) - 5600 \mbox{ kg} \times V(0.030) \approx 776 \mbox{ SWU}
\]

Another 1,000 kg of 3.5\%-enriched fuel would have to be made from other sources, for example, using natural uranium as feedstock. This would add another 4,600 SWU to the annual enrichment demand. 

In summary, using the EWLR in this way, North Korea could co-produce up to 15 kg of weapon-grade plutonium and, depending on the type of cladding, 56--116 grams of tritium per year and require about 5,400 SWU of separative work. Overall, concurrent tritium and plutonium production in the ELWR could be an attractive option if North Korea does indeed seek to maximize the inventory of both materials. However, this strategy would require developing or otherwise attaining the capacity to reprocess ceramic fuels.

It must be noted that North Korea could pursue more complex plutonium production strategies in the reactor. In particular it could use a driver-target fuel assembly design, where some fuel pins are natural or depleted uranium ``targets'' and other pins use a higher enrichment (5--10\%) to sustain criticality.\footnote{Using a driver-target fuel assembly design would enable the use of metallic uranium targets and simplify the extraction of plutonium.} Optimizing this approach would require assumptions beyond available credible information on the ELWR, such as higher UO$_2$ enrichment levels and possibly significant modifications of the core layout. Given these uncertainties, which compound the existing ones, driver-target configurations were beyond the scope of this analysis.


\bigskip

{\centering \bf Conclusion\par}%

North Korean media---from political speeches to photo releases of nuclear facilities---all but indicate the state's commitment to expanding its nuclear arsenal. There remain significant uncertainties in estimating North Korea's current inventory and production rates of weapon-grade uranium, plutonium, and tritium.

In this work, we present initial estimates of tritium production in North Korea's Experimental Light Water Reactor (ELWR) based on extensive neutronics calculations, including full reactor core simulations and various optimizations of possible fuel cycle characteristics. We focused particularly on tritium production or tritium co-production with weapon-grade plutonium. The analysis highlights, in particular, the importance of the cladding material used in the ELWR. The net tritium production rate can be about twice as high when zircaloy rather than stainless steel is used; similarly, the requirements for natural uranium and separative work drop significantly for the advanced cladding type.

Overall, our findings suggest that the ELWR can produce 48--82 grams of tritium per year. If 10 grams of tritium are allocated to each boosted nuclear weapon and half the supply is used in producing new weapons, the ELWR could enable the production of 2--4 new warheads every year or sustain a total stockpile of up to 88--150 weapons. This represents a substantially higher upper limit compared to North Korea's currently estimated fifty weapons. We also show that, in parallel with tritium production, the ELWR could produce up to 15~kg of weapon-grade plutonium per year. This would require a significantly lower burnup of the fuel, more frequent refuelings, and additional uranium and enrichment resources. Given that North Korea does not have prior experience with reprocessing of ceramic fuel, co-production of tritium and plutonium may not be a viable strategy at this time.

As North Korean officials assert that the reactor is a precursor to a dedicated commercial light-water reactor program, it is plausible that the country may further expand its nuclear power infrastructure in coming years, while taking advantage of the dual-use capabilities that these facilities may offer. 


\bigskip

{\centering \bf Acknowledgements\par}%

The authors would like to thank Siegfried Hecker for feedback on an earlier draft of this article and a conversation about the 2010 visit to the Yongbyon site, during which the North Korean hosts shared some basic design information of the ELWR. We also thank two anonymous reviewers for their comments and the participants in the 2024 School on Science and Global Security who gave feedback on the scenarios and main results discussed in this article. This work was partly supported by the Consortium for Monitoring, Technology, and Verification (MTV) under Department of Energy National Nuclear Security Administration award number DE-NA0003920.

\clearpage


\begin{table}[tb]
\centering \sf
\begin{tabular}{|r|c|c|c|}
\multicolumn{4}{c}{\textbf{\small Beginning of cycle / Initial parameters}} \\[0.5ex]
\hline
 & Clean core & Tritium core & Co-production \\
\hline \hline
Cladding material & Zircaloy-4 & Zircaloy-4 & Zircaloy-4 \\
Cladding thickness & 0.573 mm & 0.573 mm & 0.573 mm \\
\hline
Fuel enrichment & 3.50\% & 3.50\% &  3.50\% \\
\hline
Lithium-6 concentration & --- & 122 mg/kgU & 161 mg/kgU \\
Total Li load           & --- & 478 g & 631 g \\
\hline
\multicolumn{4}{c}{} \\
\multicolumn{4}{c}{\textbf{\small End of cycle / End of life of single-batch core}} \\[0.5ex]
\hline
 & Clean core & Tritium core & Co-production \\
\hline \hline
Cycle length at full power  & 804 days & 393 days & 198 days \\
Avg. discharge burnup       & 22.8 MWd/kgU & 11.2 MWd/kgU & 5.61 MWd/kgU \\
Residual U enrichment, avg. & 1.73 wt\% & 2.52 wt\% &  2.97 wt\% \\
\hline
Lithium remaining & --- & 260 g & 469 g \\
\hline
\multicolumn{4}{c}{} \\[-2ex]
\hline
T produced, decay corrected & --- & \phantom{0}95.4 g & \phantom{0}72.1 g \\
\hline
Total plutonium produced & 27.1 kg & 17.0 kg & 9.75 kg \\
Pu-239 content, avg. & 69.1 wt\% & 82.3 wt\% & 90.2 wt\% \\
\hline
\multicolumn{4}{c}{} \\
\multicolumn{4}{c}{\textbf{\small Annual requirements and production rates}} \\[0.5ex]
\hline
 & Clean core & Tritium core & Co-production \\
\hline \hline
Capacity factor & 96.4 \% & 92.9 \% & 86.8 \% \\
\hline
3.5\% uranium demand & \phantom{0}1542 kg/yr & \phantom{0}3041 kg/yr & \phantom{0}5653 kg/yr \\
Natural uranium demand & 11296 kg/yr  & 22269 kg/yr & 41402 kg/yr \\
\hline
Separative work  & 7110 SWU/yr & 14017 SWU/yr & 26060 SWU/yr \\
\hline
\multicolumn{4}{c}{} \\[-2ex]
\hline
Tritium production rate & --- & 82.2 g/yr & 115.6 g/yr \\
\hline
Plutonium production rate & 11.8 kg/yr & 14.7 kg/yr & 15.6 kg/yr \\
\hline
\end{tabular}
\caption{Fuel cycle parameters if cores use zircaloy-4 cladding. {\sf Annual rates assume 30-day outage between cycles. The separative work requirements assume natural uranium feedstock and could be significantly reduced by reusing uranium recovered from the spent fuel.}}
\label{tab:zircaloy}
\end{table}

\clearpage

\begin{table}[tb]
\centering \sf
\begin{tabular}{|r|c|c|c|}
\multicolumn{4}{c}{\textbf{\small Beginning of cycle / Initial parameters}} \\[0.5ex]
\hline
 & Clean core & Tritium core & Co-production \\
\hline \hline
Cladding material & Stainless steel & Stainless steel & Stainless steel \\
Cladding thickness & 0.400 mm & 0.400 mm & 0.400 mm \\
\hline
Fuel enrichment & 3.50\% & 3.50\% & 3.50\% \\
\hline
Lithium-6 concentration & --- & 70.0 mg/kgU & 90.0 mg/kgU \\
Total Li load           & --- & 274 g & 353 g \\
\hline
\multicolumn{4}{c}{} \\
\multicolumn{4}{c}{\textbf{\small End of cycle / End of life of single-batch core}} \\[0.5ex]
\hline
 & Clean core & Tritium core & Co-production \\
\hline \hline
Cycle length at full power  & 526 days & 291 days & 202 days \\
Avg. discharge burnup       & 14.9 MWd/kgU & 8.27 MWd/kgU & 5.72 MWd/kgU \\
Residual U enrichment, avg. & 2.24 wt\% & 2.75 wt\% & 2.96 wt\% \\
\hline
Lithium remaining & --- & 178 g & 273 g \\
\hline
\multicolumn{4}{c}{} \\[-2ex]
\hline
T produced, decay corrected & --- & \phantom{0}42.5 g & \phantom{0}35.5 g \\
\hline
Total plutonium produced & 20.8 kg & 13.3 kg & 9.82 kg \\
Pu-239 content, avg. & 77.4 wt\% & 86.2 wt\% & 90.0 wt\% \\
\hline
\multicolumn{4}{c}{} \\
\multicolumn{4}{c}{\textbf{\small Annual requirements and production rates}} \\[0.5ex]
\hline
 & Clean core & Tritium core & Co-production \\
\hline \hline
Capacity factor & 94.6\% & 90.7\% & 87.0\% \\
\hline
3.5\% uranium demand & \phantom{0}2315 kg/yr & \phantom{0}4004 kg/yr & \phantom{0}5559 kg/yr \\
Natural uranium demand & 16954 kg/yr & 29322 kg/yr & 40714 kg/yr \\
\hline
Separative work  & 10671 SWU/yr & 18457 SWU/yr & 25627 SWU/yr \\
\hline
\multicolumn{4}{c}{} \\[-2ex]
\hline
Tritium production rate & --- & 48.3 g/yr & 56.0 g/yr \\
\hline
Plutonium production rate & 13.6 kg/yr & 15.1 kg/yr  & 15.5 kg/yr \\
\hline
\end{tabular}
\caption{Fuel cycle parameters if cores use stainless steel cladding. {\sf Annual rates assume 30-day outage between cycles. The separative work requirements assume natural uranium feedstock and could be significantly reduced by reusing uranium recovered from the spent fuel.}}
\label{tab:stainless}
\end{table}

\clearpage

\vspace{-8ex} 
\begingroup 
\def\enotesize{\normalsize}
\theendnotes
\endgroup

  \end{document}